# A unified framework combining coherent compounding, harmonic imaging and angular coherence for simultaneous high-quality B-mode and tissue Doppler in ultrafast echocardiography.

Michael Mougharbel[1], Jonathan Porée[1], Stephen A. Lee[1], Paul Xing[1], Alice Wu[1], Jean-Claude Tardif[2,3] and Jean Provost[1,2]

[1]Department of Engineering Physics, Polytechnique Montréal, Montréal, QC H3T 1J4
[2]Montreal Heart Institute, Montréal, QC H1T 1C8, Canada  [3]University of Montreal, Montréal, QC H3T 1J4

**Abstract**—Various methods have been proposed to enhance image quality in ultrafast ultrasound. Coherent compounding can improve image quality using multiple steered diverging transmits when motion occurring between transmits is corrected for. Harmonic imaging, a standard technique in conventional focused echocardiography, has been adapted for ultrafast imaging, reducing clutter. Coherence-based approaches have also been shown to increase contrast in clinical settings by enhancing signals from coherent echoes and reducing clutter. Herein, we introduce a simple, unified framework that combines motion-correction, harmonic imaging, and angular-coherence, showing for the first time that their benefits can be combined in real-time. Validation was conducted through in vitro testing on a spinning disk model and in vivo on 4 volunteers. In vitro results confirmed the unified framework capability to achieve high contrast in large-motion contexts up to 17 cm/s. In vivo testing highlighted proficiency in generating images of high quality during low and high tissue velocity phases of the cardiac cycle. Specifically, during ventricular filling, the unified framework increased the gCNR from 0.47 to 0.87 when compared against coherent compounding.

## 1. INTRODUCTION

ECHOCARDIOGRAPHY is clinically used to determine the myocardium's pumping function to monitor heart failure [1], assess the structure and function of the valves [2], and diagnose pericardial diseases [3] among others. The current state-of-the-art in clinical echocardiography is based on focused ultrasound beams that form images line-by-line and by detecting the non-linear response of tissue using harmonic imaging [4] [5]. This approach leads to high image quality and to frame rates under 100 Hz in 2D [6].

To image faster phenomena, e.g., electromechanical waves [7] [8], remotely induced shear waves [9] [10], and other physiological events whose detection would improve clinical diagnosis [11] [12], higher frame rates are needed. One approach to increase frame rate is to use diverging waves [13], which can achieve up to several thousand frames per second in 2D [7] [14] [15] and 3D [15]. However, since the wavefront under such schemes is spread over a wide sector, it also degrades the point spread function and contrast. Moreover, diverging waves are attenuated both by tissue and geometrically, ultimately degrading image quality when compared against conventional imaging.

Many approaches have been proposed to improve image quality. Notably, coherently compounding several unfocused tilted transmits [14] [16], can increase image quality but at the cost of a decrease in frame rate, under the assumption of limited motion [15] [18] [19]. If motion is too large between successive transmits, received echoes become misaligned and lead to a degradation in contrast and resolution, ultimately impeding accurate diagnosis [20].

Motion compensation (MoCo) techniques have been proposed to preserve image quality throughout the cardiac cycle. Specifically, Doppler-based methods, which estimate radial velocities in-between transmits, have been used to restore image contrast and resolution[18] [19]. Beyond improving image quality, the velocity information retrieved by MoCo approaches has been used to improve high framerate speckle tracking echocardiography [21] [22] and was leveraged to design a Lagrangian beamformer which accounts for large motion when performing Ultrasound Localisation Microscopy (ULM)[21].

Harmonic imaging is a technique based on receiving the non-linear signal of tissues at multiples of the transmitted frequency and is routinely employed in conventional focused ultrasound to enhance image quality [4]. When combined with specific imaging sequences such as pulse inversion or amplitude modulation to increase the non-linear response of tissue at the expense of a reduced frame rate, harmonic imaging successfully reduces clutter noise. It can also be leveraged for ultrafast imaging and was shown to decrease clutter levels in vivo with frame rates above 150 Hz [22]. However, because of the higher frequency content and the further reduction in frame rate from harmonic sequences, ultrafast harmonic imaging is more sensitive to motion than fundamental imaging.

Coherence-based factors that rely on the degree of similarity of received wavefronts have demonstrated their ability to improve image quality as well [23]. Notably, short-lag spatial coherence, which assesses waveform similarity between pairs of receivers has shown promise with increases in CNR of 9 dB in a clinical setting [24] and was applied in thyroids, liver, kidneys, and to detect lesions in breast tissue [25]–[27]. The theory behind short-lag spatial coherence can further be extended to unfocused transmits to assess angular-coherence [28], which relies on the degree of correlation of backscattered signals which were transmitted at different angles. Angular coherence has been shown to enhance contrast by 12 dB in [29].

While all these methods have shown promise individually, they have yet to be combined in-vivo in real-time for it to be transferred to clinic. Herein, we propose a unified framework to simultaneously integrate the benefits of motion-correction, harmonic imaging and angular-coherence weighing. To design the unified framework, we adapted Doppler-based motion compensation to harmonic imaging and integrated the use of coherence information in between successive transmits to enhance the signal originating from tissue.

To validate the unified framework, an in vitro spinning disk experiment with embedded cysts was conducted. In vivo, the performance of the unified framework was assessed qualitatively and quantitatively in four participants. The proposed unified framework generated images of high contrast throughout the cardiac cycle and retrieved essential structures that would have been lost in conventional coherent compounding. and was associated with increases of gCNR of up to 0.4 in presence of large motion during ventricular filling.

## 2. THEORETICAL BACKGROUND

1) Adverse effect of motion in diverging wave imaging

In ultrasound imaging, the impulse response of the system at a given position $\vec{r}$ can be modeled as:
$$\tilde{s}(\vec{r}) = p \cdot e^{j(\vec{k}_{tx}+\vec{k}_{rx})\vec{r}} \quad (1)$$

Where $\vec{k}_{tx}$ and $\vec{k}_{rx}$ are the transmit receive wave vectors and p the pressure at that location. To improve image quality, successive transmitted beams with different tilted angles can be compounded coherently [16]. Such process can be written as:
$$\widetilde{s_C}(\vec{r}) = \sum_{m=-M/2}^{M/2} p \cdot e^{j(\vec{k}_{tx_m}+\vec{k}_{rx})\vec{r}}$$
$$\widetilde{s_C}(\vec{r}) = p \cdot e^{j(\vec{k}_{tx_0}+\vec{k}_{rx})\vec{r}} \sum_{m=-M/2}^{M/2} e^{j(m\Delta\vec{k}_{tx})\vec{r}} \quad (2)$$

Where $\vec{k}_{tx_0}$ transmit wave vector associated to a non-tilted transmit, $\Delta\vec{k}_{tx}$ the difference between successive transmit wave vector and M the number of successive transmits in a compound sequence. With the appropriate variable change, the summation can be rewritten as a discrete Fourier transform that leads to:
$$\widetilde{s_C}(\vec{r}) = p \cdot e^{j(\vec{k}_{tx_0}+\vec{k}_{rx})\vec{r}} \left[\text{sinc}\left(\frac{2M\Delta\vec{k}_{tx}\cdot\vec{r}}{2\pi}\right) * \text{III}\left(\frac{\Delta\vec{k}_{tx}\cdot\vec{r}}{2\pi}\right)\right] \quad (3)$$

Where $\text{III}(.)$ is a Dirac comb that gives the relative position of the grating lobes in transmit and the sinc function gives the width of the main lobe in transmit through compounding [18].

In presence a constant velocity $\vec{v}$ motion of the target, equation 2 becomes:
$$\widetilde{s_C}(\vec{r}) = \sum_{m=-M/2}^{M/2} p \cdot e^{j(\vec{k}_{tx_m}+\vec{k}_{rx})(\vec{r}+m.\vec{v}.T)} \quad (4)$$

Which can be rewritten, by using the previous reasoning, as:
$$\widetilde{s_C}(\vec{r}) = p \cdot e^{j(\vec{k}_{tx_0}+\vec{k}_{rx})\vec{r}} \cdot$$
$$\sum_{m=-M/2}^{M/2} e^{j(m\Delta\vec{k}_{tx})\vec{r}} e^{j(\vec{k}_{tx_0}+\vec{k}_{rx})(m.\vec{v}.T)} e^{j(m^2\Delta\vec{k}_{tx}.\vec{v}.T)} \quad (5)$$

Where T stands for the pulse repetition period of the system. The velocity of the target adds a second order term in (5) that leads to the incoherent summation of the waves and to destructive interference if the velocity is too high [17], [18].

2) Motion compensation :

To correct for the adverse effect of motion, Denarie et al. [18] proposed the use of ensemble autocorrelation over successive transmits to retrieve the Doppler shift related to motion. Ensemble autocorrelation is given by:
$$R = \sum_{m=1}^{M} \widetilde{s_m}(\vec{r})\overline{\widetilde{s_{m+1}}}(\vec{r}) \quad (6)$$

For a linearly increasing transmit sequence (i.e., $\vec{k}_{tx_m} = \vec{k}_{tx_0} + m\Delta\vec{k}_{tx}$) and a constant velocity of the target the autocorrelation becomes:
$$R = p^2 e^{-j(\vec{k}_{tx_0}+\vec{k}_{rx})\vec{v}.T} e^{-j\Delta\vec{k}_{tx}(\vec{r}+\vec{v}.T)} \sum_{m=1}^{M-1} e^{-j2m\Delta\vec{k}_{tx}.\vec{v}.T} \quad (7)$$

Which, using the same reasoning as before can be rewritten as:
$$R = p^2 e^{-j(\vec{k}_{tx_0}+\vec{k}_{rx})\vec{v}.T} e^{-j\Delta\vec{k}_{tx}(\vec{r}+\vec{v}.T)} \left[\text{sinc}\left(\frac{2M\Delta\vec{k}_{tx}.\vec{v}.T}{\pi}\right) * \text{III}\left(\frac{\Delta\vec{k}_{tx}.\vec{v}.T}{\pi}\right)\right] \quad (8)$$

Extracting the phase of the autocorrelation gives:
$$\angle R = (\vec{k}_{tx_0} + \Delta\vec{k}_{tx} + \vec{k}_{rx})\vec{v}.T + \Delta\vec{k}_{tx}.\vec{r} \quad (9)$$

The phase of the autocorrelator here is proportional to the velocity of the target plus a residual term that depends on the position of the target $\vec{r}$ and acts as an unwanted bias on the velocity estimate. Let us now write for the ascending and descending portion of the triangular sequence proposed in [17]

3(Fig 1A). For the ascending and descending portion of the sequence respectively, the transmit wave vectors are given by:

$$\vec{k}_{tx_m} = \vec{k}_{tx_0} + 2m\Delta\vec{k}_{tx}$$
$$\vec{k}_{tx_m} = \vec{k}_{tx_0} - (2m+1)\Delta\vec{k}_{tx}$$

(10)

Thus, the ascending and descending autocorrelators denoted by R1 and R2 are given by:

$$R1 = p^2 e^{-j(\vec{k}_{tx_0}+\vec{k}_{rx})\vec{v}.T} e^{-j2\Delta\vec{k}_{tx}(\vec{r}+\vec{v}.T)} \left[ sinc\left(\frac{2M\Delta\vec{k}_{tx}.\vec{v}.T}{\pi}\right) * III\left(\frac{2\Delta\vec{k}_{tx}.\vec{v}.T}{\pi}\right) \right]$$

$$R2 = p^2 e^{-j(\vec{k}_{tx_0}+\vec{k}_{rx})\vec{v}.T} e^{j2\Delta\vec{k}_{tx}(\vec{r}+\vec{v}.T)} \left[ sinc\left(\frac{2M\Delta\vec{k}_{tx}.\vec{v}.T}{\pi}\right) * III\left(\frac{2\Delta\vec{k}_{tx}.\vec{v}.T}{\pi}\right) \right]$$

(11)

The product of the two-ensemble autocorrelators then leads to the following:

$$R1.R2 = p^4 e^{-j2(\vec{k}_{tx_0}+\vec{k}_{rx})\vec{v}.T} \left[ sinc^2\left(\frac{2M\Delta\vec{k}_{tx}.\vec{v}.T}{\pi}\right) * III\left(\frac{2\Delta\vec{k}_{tx}.\vec{v}.T}{\pi}\right) \right]$$

$$\angle R1.R2 = 2(\vec{k}_{tx_0}+\vec{k}_{rx})\vec{v}.T$$

(12)

The product of these autocorrelator leads to an unbiased estimate of the Doppler velocity $v_D$ that is the projection of the velocity onto the radial axis that is defined by the transmit/receive wave vectors:

$$v_D = \frac{c}{8\pi\,Tf} \angle R1.R2$$

(13)

With c the speed of sound and f the receive frequency. This Doppler velocity estimate can then be used as in Poree et al.[17] to realign IQ images to the central transmit event of the frame determined at $(\theta,r)$ for coherent compounding:

$$\widetilde{s_{MoCo}}(\vec{r}) = \sum_{m=-M/2}^{M/2} \widetilde{s_m}(r + m.v_D.T, \theta) e^{jm\phi}$$

(14)

3) Harmonic imaging:

In second harmonic imaging the impulse response can be modeled as:

$$\tilde{s}(\vec{r}) = p e^{j(\vec{k}_{tx}+\vec{k}_{rx})\vec{r}} + p^2 e^{j(\vec{k}_{tx}+\vec{k}_{rx})\vec{r}^2}$$

(15)

Harmonic imaging through pulse inversion is done by transmitting positive and negative pulses successively. The impulse response associated to an inverted pulse transmission gives:

$$\tilde{s}(\vec{r}) = -p e^{j(\vec{k}_{tx}+\vec{k}_{rx})\vec{r}} + (-p)^2 e^{j(\vec{k}_{tx}+\vec{k}_{rx})\vec{r}^2}$$

(16)

Summing successive received signals allows for the retrieval of the second harmonic component:

$$\tilde{s}(\vec{r}) = 2p^2 e^{j2(\vec{k}_{tx}+\vec{k}_{rx})\vec{r}}$$

(17)

Thus, an adequate high-pass filter can be applied in receive to retrieve solely the harmonic component of the signal in equation 15 and 16 to perform Doppler estimation and compensation as by equations 11 to 13. Additional fundamental signal cancelation can be achieved with pulse inversion equations 17 through coherent compounding after motion compensation.

4) Angular Coherence:

With the model described above, the angular coherence Ra can be expressed as the ratio of the lag-1 to the lag-0 ensemble autocorrelation:

$$Ra = \frac{\left|\sum_{m=1}^{M-1} \widetilde{s_m}(\vec{r})\overline{\widetilde{s_{m+1}}(\vec{r})}\right|}{\left|\sum_{m=1}^{M} \widetilde{s_m}(\vec{r})\overline{\widetilde{s_m}(\vec{r})}\right|}$$

(18)

Using the models described above, after motion compensation, (18) becomes:

$$Ra = sinc\left(\frac{2M\Delta\vec{k}_{tx}.\vec{v}_{res}.T}{\pi}\right) * III\left(\frac{\Delta\vec{k}_{tx}.\vec{v}_{res}.T}{\pi}\right)$$

(19)

Where $\vec{v}_{res}$ is the residual velocity after motion compensation. Weighing $\widetilde{s_{MoCo}}(\vec{r})$ by the coherence will attenuate pixel containing large velocity residuals in addition to low signal power.

### 3. METHODS

1) Imaging sequences

A Verasonics Vantage research scanner (Verasonics Inc., Redmond, WA) and a GE-5MScD (General Electrics, Boston, MA) phased array transducer with a 2.84-MHz central frequency was used for all acquisitions. Two distinct imaging sequences with an equal number of transmits were designed to compare fundamental and harmonic imaging with pulse inversion. The parameters of the sequences are described in table I and their respective transmit schemes are shown in Fig.1

Tilt angles were arranged in a triangular configuration [17]. An angular step of 1.25° was used for both harmonic and fundamental imaging sequences. For harmonic imaging, pulse inversion was performed by transmitting positive and negative pulses successively with the same angle, leading to residual fundamental component cancelation and enhancement of the second harmonic component [4]. Diverging beams were transmitted with the whole aperture of the probe. The virtual sources positions were computed as in [18].

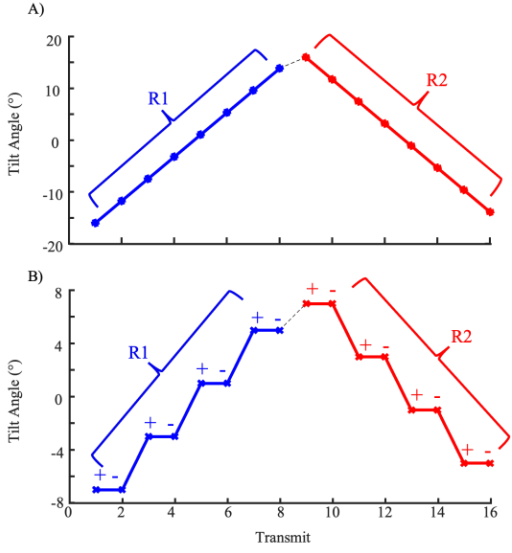

Fig. 1. Triangular transmit sequences schemes for A) Fundamental imaging and B) Harmonic imaging. For clarity, this figure shows a 16 transmit sequences, instead of the 32 transmit used in this study to acquire data. Slow-time autocorrelations were calculated in the ascending (R1) and descending stages (R2).

TABLE I
IMAGING SEQUENCE PARAMETERS

| Parameters | Fundamental | Harmonic |
|---|---|---|
| Transmit frequency | 2.98 MHz | 1.89 MHz |
| Receive frequency | 2.98 MHz | 3.78 MHz |
| Number of cycles | 3 | |
| Pulse Repetition Frequency (PRF) | 4700 Hz | |
| Frame Rate | 140 Hz | |
| Number of transmitted diverging waves | 32 | |
| Imaging sector width | 75° | |

2) Beamforming and post-processing

Beamforming and compounding: The received radiofrequency signals (RFs) were demodulated at 2.98 MHz and 3.78 MHz for fundamental and harmonic imaging respectively. Beamforming was performed using an In-Phase and Quadrature (IQ) delay-and-sum algorithm [29] with a speed of sound of 1540 m/s. The reconstruction grid $(\theta, r)$ had an angular resolution of $\Delta\theta = 0.4°$ and a radial resolution of $\Delta r = \frac{\lambda}{4}$ where $\lambda$ is the wavelength computed at the central frequency of the transducer (2.84 MHz). After beamforming, 32 diverging waves were compounded such as in [16] for both fundamental and harmonic imaging.

Motion compensation implementation: MoCo was implemented as described in section II. B: Ascending R1 and descending R2 autocorrelator were evaluated using equations 7, which yields:

$$R1 = \sum_{m=1}^{M/2} \widetilde{s_m}(\vec{r})\overline{\widetilde{s_{m+1}}}(\vec{r})$$

$$R2 = \sum_{m=M/2}^{M} \widetilde{s_m}(\vec{r})\overline{\widetilde{s_{m+1}}}(\vec{r})$$

(20)

After which equation 13 and 14 were used to coherently compound motion corrected images associated to individual transmits. When correcting for motion, the theoretical maximal speed $V_{DMax}$ that we can correct for to respect the Nyquist limit is given by: $V_{DMax} = \frac{PRF\,c}{8f}$ . which yields 30 cm/s for fundamental (f=2.98 MHz) and 24 cm/s for harmonic imaging (f=3.78 MHz) respectively.

Angular coherence implementation: The normalized lag-one and lag-two autocorrelator denoted by $w_F$ and $w_H$ in equation 22 and 23 for fundamental and harmonic imaging respectively, were computed to indicate the level of angular coherence between successive tilted transmits.

$$w_F = \frac{|\sum_{m=1}^{M-1} \widetilde{s_{mc}}\overline{\widetilde{s_{mc+1}}}|}{|\sum_{m=1}^{M-1} \widetilde{s_{mc}}\overline{\widetilde{s_{mc}}}|} \quad (22)$$

$$w_H = \frac{|\sum_{m=1}^{M-2} \widetilde{s_{mc}}\overline{\widetilde{s_{mc+2}}}|}{|\sum_{m=1}^{M-2} \widetilde{s_{mc}}\overline{\widetilde{s_{mc}}}|} \quad (23)$$

Then, motion-compensated frames were weighted with the autocorrelators amplitude for both fundamental and harmonic imaging.

$$\widetilde{s_{WF}}(\theta, r) = w_F \cdot \widetilde{s_{MoCoF}} \quad (24)$$

$$\widetilde{s_{WH}}(\theta, r) = w_H \cdot \widetilde{s_{MoCoH}} \quad (25)$$

Where $\widetilde{s_{MoCoF}}$ and $\widetilde{s_{MoCoH}}$ are the motion-compensated frames for fundamental and harmonic imaging, respectively. These weights indicate the level of angular coherence and amplify signals originating from strongly correlated structures such as tissue. Lag-2 is used in harmonic imaging to account for pulse inversion.

The pipeline (Fig 2) is applied to either fundamental or harmonic data. Six combinations (2 sequences x 3 processing methods) are compared to evaluate the influence of harmonic imaging, motion compensation, and angular coherence, which together constitute what we refer herein to a unified framework. The experimental design includes: (1) Standard compounding with only coherent compounding, (2) MoCo applied on IQ signals before coherent compounding, and (3) Coherence-weighted MoCo.

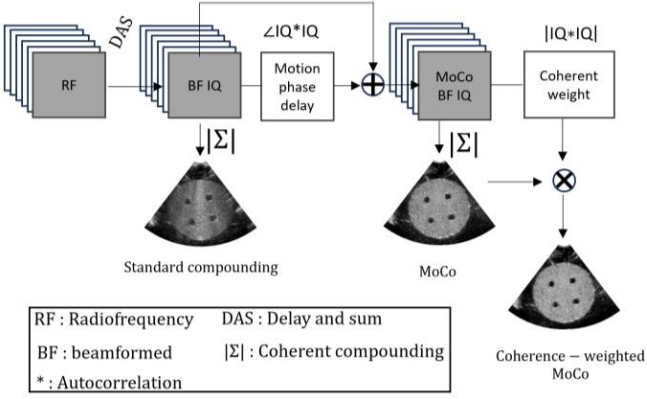

Fig. 2 Pipeline to generate coherence-weighted motion-compensated images. — From left to right:. 1) RFs data are received. 2) RFs data are beamformed to get complex images (BF IQ), coherent compounding of the BF IQ generates a B-mode frames. 3) Slow-time autocorrelations yields phase delays induced by tissue motion. 4) The phase delays are used to realign the beamformed complex images and get motion-compensated beamformed complex frames (MoCo BF IQ), coherent compounding of the MoCo BF IQ generates motion-compensated B-mode images 5) Amplitude of the slow time autocorrelation returns a coherent image highlighting strongly correlated structures along the slow time. The multiplication of the B-mode resulting from (MoCo BF IQ) and the coherence image yields coherence-weighted B-mode images.

## 4. Experiments

### 1) In vitro

In vitro experiments were performed in a 11-cm diameter spinning agar disk (94% Deionized water, 5% Agar, 1% Benzallkonium chloride). Four cysts of 1.25 cm diameter were embedded 2.75 cm away from the center. Radial velocities of cysts varied from 0 to 26 cm/s. To assess the effect of motion on image quality the generalized Contrast to Noise Ratio (gCNR) [30] was used. The cysts were chosen to be the region of interest and annuluses of the same area surrounding the cyst were chosen as the background region. The gCNR was computed when the disk had maximal and minimal radial velocities, respectively (see Fig.4).

The estimated Doppler velocities of the spinning disk $\widehat{v_D}$ (see equation 13), were compared to the true velocities at which the disk was spinning. The disk's spinning speed was adjusted using a potentiometer that controlled a connected DC motor and the true velocities were measured using a tachometer, which recorded the disk's rotations per minute. To quantify the performance of the estimated Doppler velocities, the normalized root mean square error (NRMSE) was calculated as:

$$\text{NRMSE} = \frac{\sqrt{\frac{1}{N}\sum_i^N (v_{Di} - \widehat{v_{Di}})^2}}{r_{cyst} \cdot \omega}$$

(26)

Where $v_{Di}$, and $\widehat{v_{Di}}$ are the true and Doppler estimated radial velocities respectively at pixel i, N the total number of pixels in the ROI, $\omega$ the known angular velocity and $r_{cyst}$ the distance between cysts and the disk center (2.75 cm in this case)

### 2) In-vivo

Transthoracic cardiac acquisitions were performed in 4 volunteers with no known cardiac conditions. Acoustic outputs for both fundamental and harmonic sequences were measured using a calibrated hydrophone (Acertara, Longmont, CO), to ensure the ultrasonic sequences complied with the Food and Drug Administration (FDA) requirements for cardiac imaging (510 k Track 3, FDA). The Mechanical Index (MI) for fundamental and harmonic imaging was measured at 0.18 and 0.25 respectively (FDA limit set at 1.9), while the Spatial peak temporal average intensity had a value of 38 and 68 mW/cm$^2$ for fundamental and harmonic imaging respectively (FDA limit set at 430 mW/cm$^2$). These acoustic outputs were calculated with a derating factor of $0.3\ \text{dB.cm}^{-1}.\text{MHz}^{-1}$ to account for tissue attenuation.

For all acquisitions, the apical 4-chamber view, parasternal short axis (SAX), and parasternal long axis (PSLA) were performed to assess image quality. When switching from the harmonic to the fundamental imaging sequence, the position of the probe was maintained. The protocol for in vivo acquisition was reviewed and approved by Polytechnique Montréal's ethic board (CER-2122-54-D) and volunteers signed an informed consent form to participate in the study.

For quantitative assessment, the apical 4-chamber view was chosen to perform gCNR measurements as the septum is most visible in this view and is convenient to select regions of interest. Hence, for each volunteer, two regions of interest in the septum were chosen for gCNR calculations, with the cavities serving as the background region. These measurements were performed during diastasis, (i.e, minimal tissue motion) and rapid ventricular filling (i.e, maximal tissue motion).

## 5. Results

### 1) In vitro results:

The images of the spinning disk, shown in Fig. 3, demonstrate that when standard compounding is performed, the contrast of the cysts degrades as soon as minor motion is introduced (i.e., cysts moving at 5 cm/s). As the speed of the disk increases to reach 24 cm/s at cyst level, cysts become indistinguishable. The observation holds true for both fundamental and harmonic imaging which yield similar image quality.

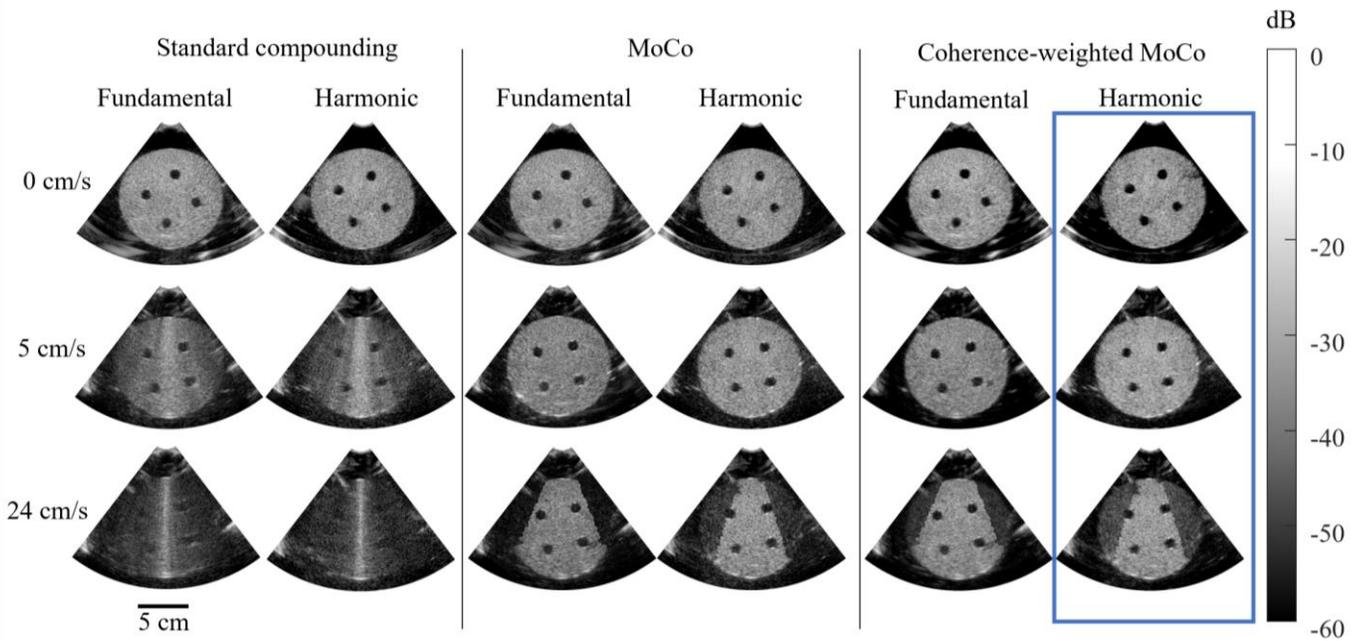

Fig. 2 Compounded images of the in vitro disk imaged with fundamental and harmonic imaging at different speeds (0 cm/s, 5cm/s and 24 cm/s) processed with standard compounding, MoCo and coherence-weighted MoCo. The blue rectangle highlights the results obtained with the unified framework

When MoCo is applied, it successfully generates images with contrast quality comparable to those produced under motionless conditions if the speed applied on the disk remains under the Nyquist limit (30 and 24 cm/s for fundamental and harmonic, respectively). Above these velocities, decorrelation appears due to Doppler aliasing as observed in the bottom right panel. In the case of coherence-weighted MoCo, we find that the contrast of the images closely resembles those generated when only MoCo is applied across all speed conditions for both fundamental imaging and the unified framework, which uses harmonic imaging.

Figure 4 quantitatively evaluates the contrast of disk images across the six different combinations. For the south cyst exhibiting maximal azimuthal and minimal radial motion (Fig. 4A), when standard compounding is performed, contrast decreased gradually as motion increases. The gCNR drops from 0.77 in the standard method to between 0.42 and 0.63 for velocities above 15 cm/s in both imaging methods. With MoCo applied, gCNR stays between 0.73 and 0.8 for speeds from 0 to 26 cm/s in fundamental and harmonic imaging, exhibiting similar performance. Coherence-weighted MoCo follows similar trends with a small improvement in gCNR for both imaging types. Specifically, there's an average 4% increase in gCNR for speeds from 0 to 26 cm/s compared to only MoCo, in both imaging methods.

For the northwest cyst exhibiting maximal radial velocity (Fig.4B), for standard compounding, contrast decreases rapidly as soon as motion is introduced and worsens as speed increases. Indeed, gCNR dropped from 0.80 when there is no motion to range between 0.05 and 0.25 for velocities greater than 15 cm/s for harmonic and fundamental imaging. When MoCo is applied, contrast is recovered and gCNR is maintained at an average of $0.84 \pm 0.01$ for fundamental imaging and $0.81 \pm 0.01$ for harmonic imaging for speeds ranging from 0 to 17 cm/s. Near the Nyquist limit at 17 cm/s, an expected decrease in gCNR is observed for harmonic imaging as decorrelation appears faster in harmonic imaging than fundamental imaging. For coherence-weighted MoCo, the curves' trends are similar to the only MoCo condition with a small improvement in gCNR. In fact, for speeds ranging from 0 to 17 cm/s there is an average increase of 2% and 2.5% when compared to only MoCo for fundamental and harmonic imaging respectively.

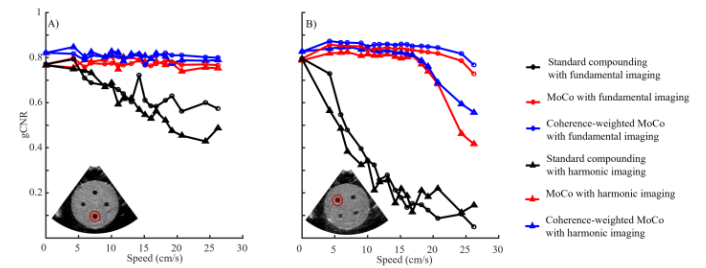

Fig. 4: Effect of speed on cyst contrast A) at maximal azimuthal velocity and B). at maximal radial velocity —Harmonic imaging (triangles), fundamental imaging, (circles) were tested with standard compounding (black), MoCo (red) and coherence-weighted MoCo (blue). The innermost red circle delineates the cyst, serving as the region of interest, while the annulus formed by the two concentric red circles functions as the background for the calculation of the gCNR

Fig. 5 shows the NRMSE curve, which quantifies the error on the Doppler velocity estimated at cyst levels (Fig. 5A). Fig. 5B shows the estimated Doppler velocity fields rendered when compared with the theoretical radial speed. Specifically, when the cysts are rotating at a speed below the Nyquist limit (e.g., 5 cm/s), the estimated Doppler velocity field matches the theorical velocity. However, when cysts are moving at 24 cm/s (i.e., the maximal velocity that can be estimated when combining MoCo with harmonic imaging), discrepancies start to appear in harmonic imaging whereas Doppler velocities estimates remain accurate for fundamental imaging. These findings concur with the NRMSE curve. Indeed, when speeds range between 4 cm/s and 21 cm/s, the NRMSE remains below 12% for fundamental and harmonic imaging. However, NRMSE starts to increase considerably for harmonic imaging, reaching 14% and 28% when velocities reached 24 cm/s and 26 cm/s respectively.

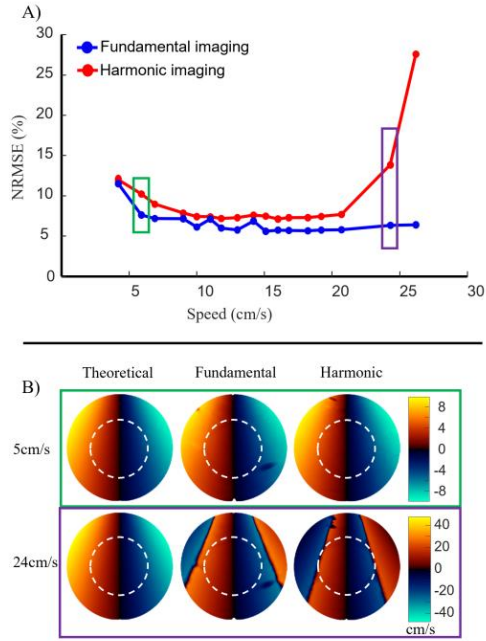

Fig. 5: Effect of speed on Doppler estimates— A) Doppler NRMSE vs. speed when the disk was imaged with fundamental imaging (blue) and harmonic imaging (red). The green and purple box refer to the datapoints for which velocity profiles were displayed. B) Theoretical and estimated velocity profiles, the doted circles show the area for which the NRMSE was computed as their radius correspond to the cyst distance

2) In vivo results:

Fig. 6A and Fig. 6B qualitatively assess B-mode images of the apical four chamber view under the six different combinations to show the contribution of MoCo, angular coherence and harmonic imaging in vivo.

Additionally, images were assessed during diastasis (i.e., when tissue motion is at its lowest) and ventricular filling to determine if applying the unified framework during periods of large tissue motion enables the retrieval of images with quality comparable to those captured during periods of minimal motion. The data shown in (Fig. 7, Supplementary Video 1) are indicative of the quality obtained across all volunteers of this study.

During diastasis (Fig. 6A), the images generated with standard compounding and MoCo exhibit comparable image quality. Moreover, coherence-weighted MoCo increases contrast of B-modes for both fundamental and harmonic imaging. When compared to fundamental imaging, harmonic imaging generated better contrasted images as the septum is more discernible from the ventricles and clutter is reduced at the atrium levels. During ventricular filling (i.e., when tissue motion is at its highest) (Fig. 6B), decorrelation leads to blurring of the septum, rendering it undistinguishable in the B-mode images generated by standard compounding for both fundamental and harmonic imaging. The application of MoCo resulted in an improvement of image quality characterized by the retrieval and well-defined depiction of the septum. Under this processing, image quality is comparable to the one generated by coherent compounding during diastasis for fundamental and harmonic imaging, respectively. In the coherence-weighted MoCo approach, contrast is further increased when compared to standard MoCo as speckle is less present in the cavities and the septum signal is enhanced.

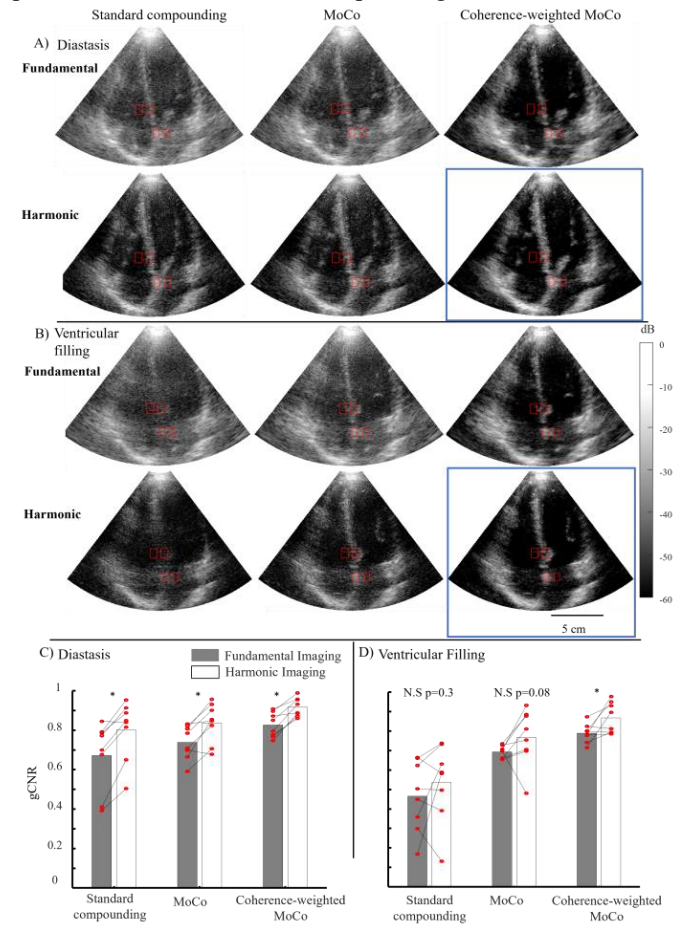

Fig. 6: In vivo transthoracic B-mode images in a human heart A) during diastasis and B) during ventricular filling. The red rectangles surrounding tissue highlight the region of interest, while the rectangles in the cavities highlight the background used for the gCNR calculation. The blue rectangle highlights the results obtained with the unified framework. Panel C) and D) show average gCNR across volunteers during diastasis and ventricular filling respectively. The lines connecting the red circles link the corresponding ROIs when processed with fundamental and harmonic imaging. A star above histograms indicates statistical significance (p≤0.05), and p-values are provided when the paired t-test reveals a non-significant (N.S) difference.

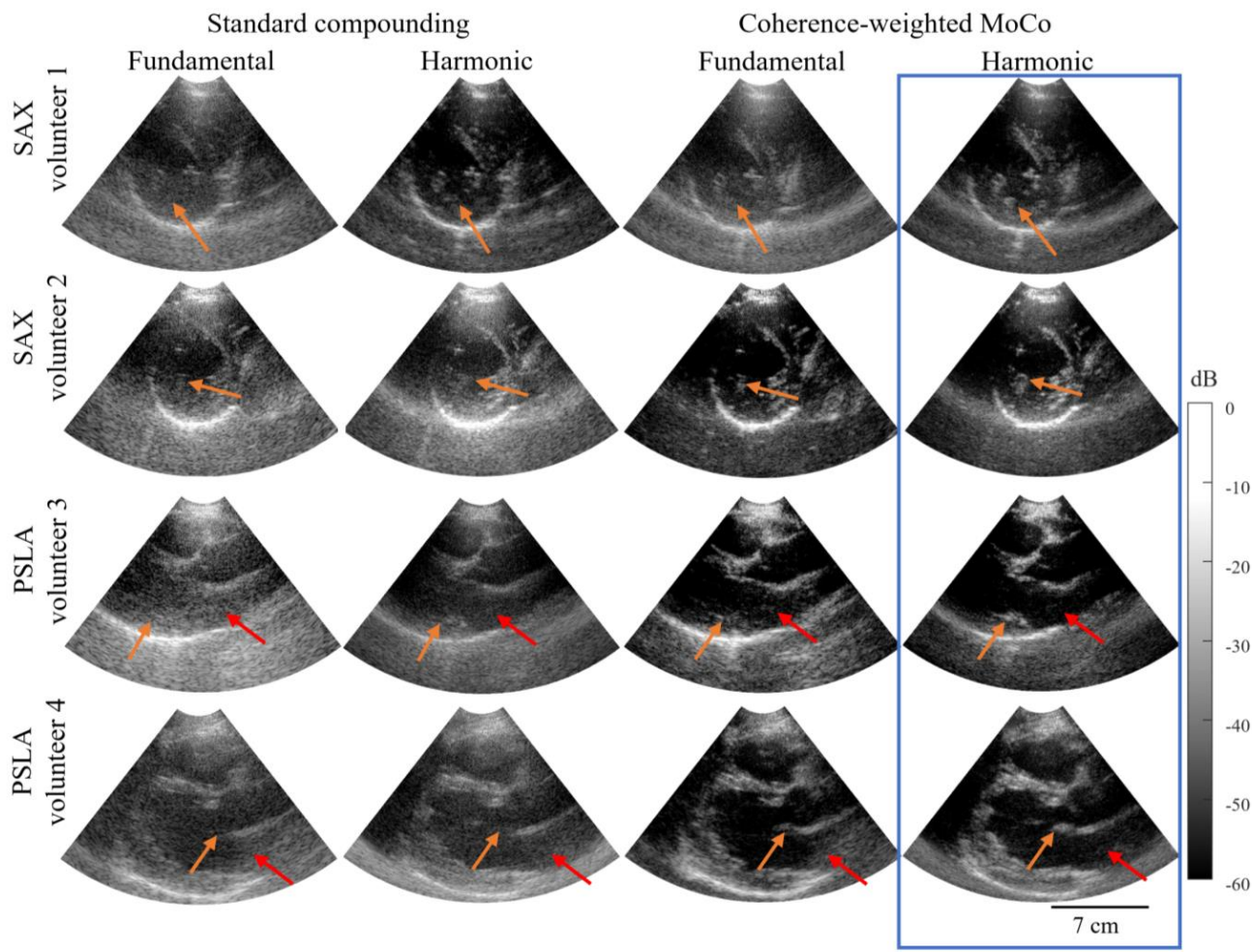

Fig. 7 In vivo transthoracic B-mode images in the human heart of four volunteers during ventricular filling. The heart was imaged with fundamental and harmonic imaging processed with standard compounding and coherence-weighted MoCo. The blue rectangle highlights the results obtained with the unified framework

Fig. 6C and Fig. 6D quantitatively evaluate the contrast of B-mode images and highlight the positive effect of each of the unified framework components. During diastasis (Fig. 6C), when the gCNR was averaged on the ROIs across volunteers, their values gradually increased when MoCo and then coherentce weighted MoCo were applied for both fundamental and harmonic imaging. Specifically, for fundamental imaging, the average gCNR value was of 0.67 for standard compounding, 0.74 when MoCo was applied and 0.83 in the coherence-weighted MoCo case. For harmonic imaging, the average gCNR value was 0.80 for standard compounding, 0.83 when MoCo was applied, and 0.92 in the coherence-weighted MoCo case. The paired t-test showed that harmonic imaging significantly increased the gCNR of the images when compared to fundamental imaging.

During rapid ventricular filling (Fig. 6D), applying MoCo improved gCNR when compared to standard compounding. In fact, gCNR values achieved when applying only MoCo during this phase are comparable to those obtained during diastasis for both fundamental and harmonic imaging. Moreover, gCNR values are increased when adding the angular coherence information on motion-compensated frames. Specifically, for fundamental imaging the average gCNR value was of 0.47 for standard compounding, 0.69 when MoCo was applied and 0.79 in the coherence-weighted MoCo case. For harmonic imaging, the average gCNR value was 0.53 with standard compounding, 0.77 when MoCo was applied and 0.87 in the coherence-weighted MoCo case. In all cases, the average values yielded by harmonic imaging are higher when compared to fundamental imaging and is significant for coherence-weighted MoCo according to the paired t-test.

Fig. 7 and Supplementary videos 2 and 3 shows PSLA views and SAX views during ventricular filling in two volunteers. and illustrates that coherence-weighted MoCo qualitatively improves image quality when compared to regular compounding, especially when it's combined with harmonic imaging. Indeed, orange arrows show structures that are better retrieved with the unified framework. Specifically, the papillary muscle in the SAX view and the mitral valve in the PSLA view are better retrieved when weighted motion compensation is applied, especially in the case of harmonic imaging. As for the red arrows, they show regions for which clutter is less visible when using coherence-weighted MoCo, especially when it's combined with harmonic imaging. The positive effect of the unified framework can also be seen in real time (Supplementary video 4)

## 6. Discussion

A unified framework combining the benefits of motion correction, harmonic imaging, and angular coherence in high frame rate echocardiography was proposed and investigated in this study. It was validated in vitro in a spinning disk and in 4 volunteers in vivo and showed its capability to yield high contrasted images throughout the cardiac cycle. A key feature of the proposed approach is that speed estimation is performed between each transmit, which lifts the typical trade-off between frame-rate and the number of compounded angles.

When assessing the contribution of each component of the unified framework on the spinning disk, we showed that MoCo allows to retrieve the structures within the disk (Fig. 3). Regarding the angular coherence, gCNR curves in Fig. 4 showed that adding its information increases the contrast as it enhances signal from correlated structures in between transmits and reduces the speckle noise. As for harmonic imaging, its performance was comparable to fundamental imaging, explained by the ideal conditions of the in vitro medium, free of structures that would typically contribute to clutter in an in vivo setting. Such conditions do not fully leverage the benefits of harmonic imaging, which minimizes imaging artefacts like clutter and relying non-linear characteristics of tissues to enhance signal quality. Hence, the in vitro set up can be improved by adding an aberrating layer on the disk

In vivo, the unified framework was assessed during different phases of the cardiac cycle characterized by their maximal and minimal tissue motion (i.e., ventricular filling and diastasis respectively) and generated images of high contrast with a gCNR above 0.85 under both conditions outperforming other combinations. The contribution of each component of the unified framework can be assessed in vivo. During ventricular filling, MoCo allowed to retrieve the septum and valves that were otherwise blurred in standard compounding. Indeed, coherent compounding requires motion smaller than $\lambda/6$ between the first and last compounded diverging wave transmission to avoid image degradation [15], which in echocardiography is typically not the case. Angular coherence information to weight the motion-compensated frames amplified the visibility of the septum and the valves while reducing speckle noise within the cavity, ultimately increasing contrast. As for harmonic imaging, it enhanced images quality, during diastasis and ventricular filling by removing clutter, which is consistent with the findings in [23] and is supported by the conducted statistical analysis.

The unified framework has successfully been applied in all the volunteers and different standard views including SAX, PSLA and apical 4-chamber views showing its robustness. Indeed, acquisitions during ventricular filling revealed that the application of the unified framework retrieved important structures used for diagnosis such as the papillary muscle, the mitral valve and the septum which were otherwise lost if only coherent compounding was performed. In vivo experiments were also performed using low MI, well-suited for adaptation to contrast-enhanced ultrasound or Ultrasound Localization Microscopy that rely on imaging contrast agents with low MI to avoid microbubble destruction.

The unified framework only corrects for radial motion and is not fully two dimensional since it cannot correct for azimuthal motion. Other methods [32], [33] have been proposed for two dimensional corrections, but are computationally expensive compared to the MoCo of the unified framework which can be cast as a simple autocorrelation operation followed by linear interpolation (See appendix A). Moreover, Fig. 4 A shows that the azimuthal motion has a lesser effect on contrast, even at high velocities. A further limitation of the unified framework is that its MoCo speed estimation is constrained by the Nyquist limit. Moreover, to address scenarios, e.g., where stress conditions may elevate tissue speeds beyond the unified framework's maximum correction capability, dealiasing techniques can be employed [34], [35]. Finally, the unified framework does not account for out of plane motion so a 3D implementation would be of interest.

## 7. Conclusion

In this work, the feasibility of combined motion compensation, harmonic imaging, and angular coherence information under a unified framework was investigated. The unified framework was validated in vitro and in vivo, and our findings showed that it generates high quality images even during phases of the cardiac cycle characterized by important motion. The proposed framework is straightforward to implement in real time which paves the way for real time high frame rate high contrast echocardiography.

## Appendix A

This code of the unified framework allows an easy implementation and demonstrate its simplicity.

```matlab
function [s_moco, vD, agl_corr] = Unified_MoCo(s,O,R,c,f,PRP)

% Input:
% s : beamformed iq signal prior compounding
% [O,R] : Polar Grid (use meshgrid(o,r))
% c : speed of sound
% f : central frequency of the received signals
% PRP : Pulse Repetition Period (1/PRF)

ntx = size(s,3);
% Doppler Ensemble autocorrelation
R1  = sum(s(:,:,1:ntx/2-1).*conj(s(:,:,2:ntx/2)),3);
R2  = sum(s(:,:,ntx/2:end-1).*conj(s(:,:,ntx/2+1:end)),3);
Phi = angle(R1.*R2)/2;
vD  = c/(4*pi*f*PRP)*Phi;

% Motion Correction
dR  = vD.*PRP;
for n = 1:ntx
    sc(:,:,n) = interp2(O,R,...
        s(:,:,n),...
        O,R+(n-ntx/2).*dR,'linear');
    sc(:,:,n) = sc(:,:,n).*exp(j*(n-ntx/2).*Phi);
end
s_moco = sum(sc,3);

% Angular Coherence
R1 = sum(s(:,:,1:ntx-1).*conj(s(:,:,2:ntx)),3);
R0 = sum(s(:,:,1:ntx).*conj(s(:,:,1:ntx)),3);

agl_corr = abs( R1 )./abs( R0 );

return
```